\documentclass[superscriptaddress,reprint,secnumarabic,amssymb, nobibnotes, apl]{revtex4-1}

\DeclareMathAlphabet{\mathpzc}{OT1}{pzc}{m}{it}

\usepackage{graphics}
\usepackage{epsfig}
\usepackage{epstopdf}
\usepackage{subfigure}
\usepackage{amsmath}
\usepackage{subfigure}
\usepackage{float}
\usepackage{enumerate}
\usepackage{array}
\usepackage{pifont}
\usepackage[11pt]{moresize}
\usepackage{datetime}

\usepackage{mathrsfs}
\newcolumntype{C}[1]{>{\centering\let\newline\\\arraybackslash\hspace{0pt}}m{#1}}
\newcolumntype{N}{@{}m{0pt}@{}}

\begin{document}

\title{Thermal graphene metamaterials and epsilon-near-zero high temperature plasmonics}
\author{Sarang Pendharker}
\author{Huan Hu}%
\affiliation{ 
Department of  Electrical and Computer Engineering University of Alberta, Edmonton, AB T6G 2V4, Canada
}%

\author{Sean Molesky}
\affiliation{ 
Department of  Electrical and Computer Engineering University of Alberta, Edmonton, AB T6G 2V4, Canada
}%
\author{Ryan Starko-Bowes}
\affiliation{ 
Department of  Electrical and Computer Engineering University of Alberta, Edmonton, AB T6G 2V4, Canada
}%
\author{Zohreh Poursoti}
\affiliation{ 
Department of  Electrical and Computer Engineering University of Alberta, Edmonton, AB T6G 2V4, Canada
}%

\author{Sandipan Pramanik}
\affiliation{ 
Department of  Electrical and Computer Engineering University of Alberta, Edmonton, AB T6G 2V4, Canada
}%

\author{Neda Nazemifard}
\affiliation{ 
Department of  Chemical and Materials Engineering University of Alberta, Edmonton, AB T6G 2V4, Canada%
}
\author{Robert Fedosejevs}
\affiliation{ 
Department of  Electrical and Computer Engineering University of Alberta, Edmonton, AB T6G 2V4, Canada
}%

\author{Thomas Thundat}
\affiliation{ 
Department of  Chemical and Materials Engineering University of Alberta, Edmonton, AB T6G 2V4, Canada%
}
\author{Zubin Jacob}
\affiliation{ 
Department of  Electrical and Computer Engineering University of Alberta, Edmonton, AB T6G 2V4, Canada%
}
\affiliation{
Birck Nanotechnology Center, School of Electrical and Computer Engineering,
Purdue University, West Lafayette, IN 47906, USA%
}
\email{zjacob@ualberta.ca}

\begin{abstract}
The key feature of a thermophotovoltaic (TPV) emitter is the enhancement of thermal emission corresponding to energies just above the bandgap of the absorbing photovoltaic cell and simultaneous suppression of thermal emission below the bandgap. We show here that a single layer plasmonic coating can perform this task with high efficiency. Our key design principle involves tuning the epsilon-near-zero frequency (plasma frequency) of the metal acting as a thermal emitter to the electronic bandgap of the semiconducting cell. This approach utilizes the change in reflectivity of a metal near its plasma frequency (epsilon-near-zero frequency) to lead to spectrally selective thermal emission and can be adapted to large area coatings using high temperature plasmonic materials. We provide a detailed analysis of the spectral and angular performance of high temperature plasmonic coatings as TPV emitters. We show the potential of such high temperature plasmonic thermal emitter coatings (p-TECs) for narrowband near-field thermal emission. We also show the enhancement of near-surface energy density in graphene-multilayer thermal metamaterials due to a topological transition at an effective epsilon-near-zero frequency. This opens up spectrally selective thermal emission from graphene multilayers in the infrared frequency regime. Our design paves the way for the development of single layer p-TECs and graphene multilayers for spectrally selective radiative heat transfer applications.

\end{abstract}

\maketitle

\section{Introduction}

\begin{figure*}
\includegraphics[width=1\linewidth]{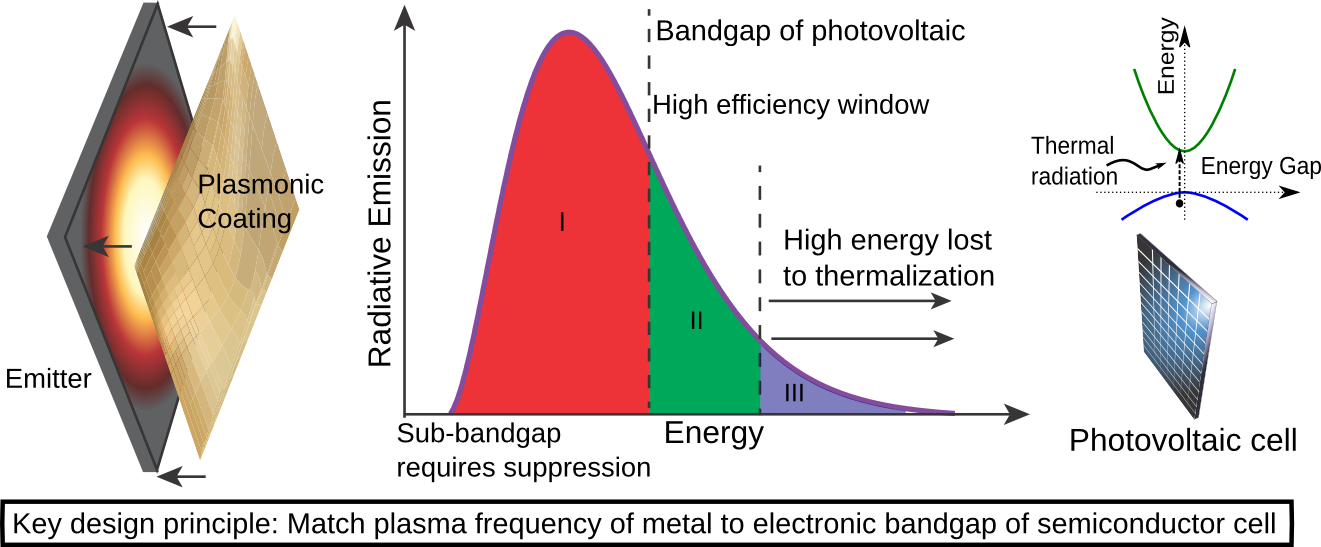}
\caption{\label{fig:Fig1} Efficient TPV energy conversion requires two critical features: (1) Suppression of thermal radiation with energy less than the band-gap of the  PV cell. (2) Enhancement of thermal radiation within the high efficiency window of energies which lies directly above the PV cell band-gap. Both of these goals can be accomplished by tuning the plasma frequency (epsilon-near-zero frequency) of a high temperature plasmonic thermal emitter coating (p-TEC).}
\end{figure*}

Thermophotovoltaic systems promise to provide high efficiency energy conversion from heat to electricity by spectrally matching the thermal emission from an emitter to the bandgap of a conventional single-junction photovoltaic cell \cite{bauer2011thermophotovoltaics,ilic2012overcoming,park2008performance,nefzaoui2012selective,molesky2015ideal}. This field of research has seen a resurgence in recent years due to parallel developments in the fields of nanoscale thermal engineering \cite{nagasaka2008zhuomin}, nanofabrication \cite{fleming2003three,yeng2012enabling} and high temperature material science \cite{guler2014refractory,molesky2013high,guo2014thermal,nagpal2008efficient}.  An important aspect of the thermophotovoltaic system is the design of the emitter, which has to suppress the thermal emission of sub-bandgap photons as they cannot be converted to electrical output \cite{fleming2003three,yeng2012enabling} and simultaneously enhance thermal emission just above the cell bandgap. Note that sub-bandgap blackbody photons are the primary reason for decrease in the efficiency of energy conversion (see Fig.~\ref{fig:Fig1}).  One necessary requirement for an emitter is robust optical response and thermal/structural stability of the emitter which has to withstand high temperatures \cite{bauer2011thermophotovoltaics}.

Significant recent advances have shown the potential of photonic crystals \cite{fleming2003three,yeng2012enabling,nagpal2008efficient,rephaeli2009absorber,narayanaswamy2005thermal, celanovic20041d} , thin films \cite{guo2014thermal,wang2015tunneling,song2015enhancement,basu2011maximum, edalatpour2013size, dimatteo2001enhanced, narayanaswamy2003surface,park2008performance, joulain2008near,nefedov2011giant,tong2015thin}, perfect absorbers \cite{tong2015thin}, thermal metamaterials \cite{dyachenko2016controlling,molesky2013high,mason2011strong,liu2011taming,wu2012metamaterial}, tungsten metasurfaces \cite{zhao2013thermophotovoltaic,neuner2013efficient,wu2012metamaterial}, graphene \cite{messina2012graphene, svetovoy2014graphene}, surface waves \cite{ben2013controlling,joulain2005surface} and other nanophotonic structures \cite{ilic2012overcoming,deng2015broadband} to engineer the thermal emission \cite{drevillon2011far,lee2007coherent} and achieve spectrally selective emitters \cite{drevillon2011far, wang2009spatial,lee2007coherent}. Simultaneously, work has progressed to advance the critical challenge involving the absorber/TPV cell design. 

Here, we build on our previous work on the first suggestion of high temperature plasmonics \cite{guo2014thermal,molesky2013high} and experimental demonstration of high temperature thermal metamaterials \cite{dyachenko2016controlling}. In this paper, we provide an alternative in selective emitter design employing only a single layer of plasmonic thermal emitter coating. This design holds the significant advantage of not requiring 2D patterning to achieve the desired optical response. Along with ease of large area fabrication it also opens up the possibility of high temperature stable operation since 2D texturing often reduces the melting point of metals. The primary objective of thermal suppression of sub-bandgap photons can be achieved by tuning the epsilon-near-zero (ENZ) frequency (also known as plasma frequency) of a metal to the bandgap of the cell. Thus natural material properties provide the reflectivity change desired for the thermal emitter as opposed to structural bandgap effects as in photonic crystals. We also show that our design is useful for engineering narrowband near-field thermal emission paving the way for a unified platform for both far-field and near-field thermal emitters. We also provide detailed performance analysis for our thin film designs. It should be noted that polar dielectric thin films (eg: silicon carbide) only show phonon-polaritonic resonances and epsilon-near-zero regime only in the mid-infrared frequency ranges and not in the near-infrared range conducive for high temperature thermophotovoltaic applications. 

One important application of selective thermal emitters is the ability to modulate thermal emission along with spectrally selective nature \cite{brar2015electronic,van2011fast,fang2013active,freitag2010thermal} . Graphene presents an ideal platform to achieve this effect due to the strong electrical tunability of its optical absorption properties \cite{thongrattanasiri2012complete} . Graphene is expected to find wide applications in next generation electronic devices due to its good electrical \cite{bolotin2008ultrahigh} and thermal properties \cite{ghosh2008extremely} and strong light matter interactions \cite{koppens2011graphene}. A substrate coated with a graphene layer has been reported to have highly directive far-field thermal emission, enhanced light absorption \cite{pu2013strong} and enhanced near-field radiation transfer \cite{ilic2012near,lim2013near}. An important emerging platform are the multilayer graphene based metamaterials \cite{iorsh2013hyperbolic,zhang2014tunable,sreekanth2013negative}, which have been reported to exhibit tunable absorption and hyperbolic dispersion \cite{othman2013graphene,linder2016graphene,chang2016realization}. 

In this paper, we report a sharp suppression in the thermal emissivity of graphene-multilayer structure, and show that it can be controlled by tuning the topological transition in the graphene metamaterial. We also report a sharp enhancement in the near-field thermal emission in graphene-multilayers due to topological transitions. These results forms a crucial step in the realization of thermal energy scavenging technologies in future graphene based electronic devices. We note that the opportunity to electrically tune topological transitions can be an important design principle for spectrally selective thermal modulation.

\section{High temperature plasmonic thermal emitter coating (p-TEC)}

\begin{figure}
\includegraphics[width=0.5\textwidth]{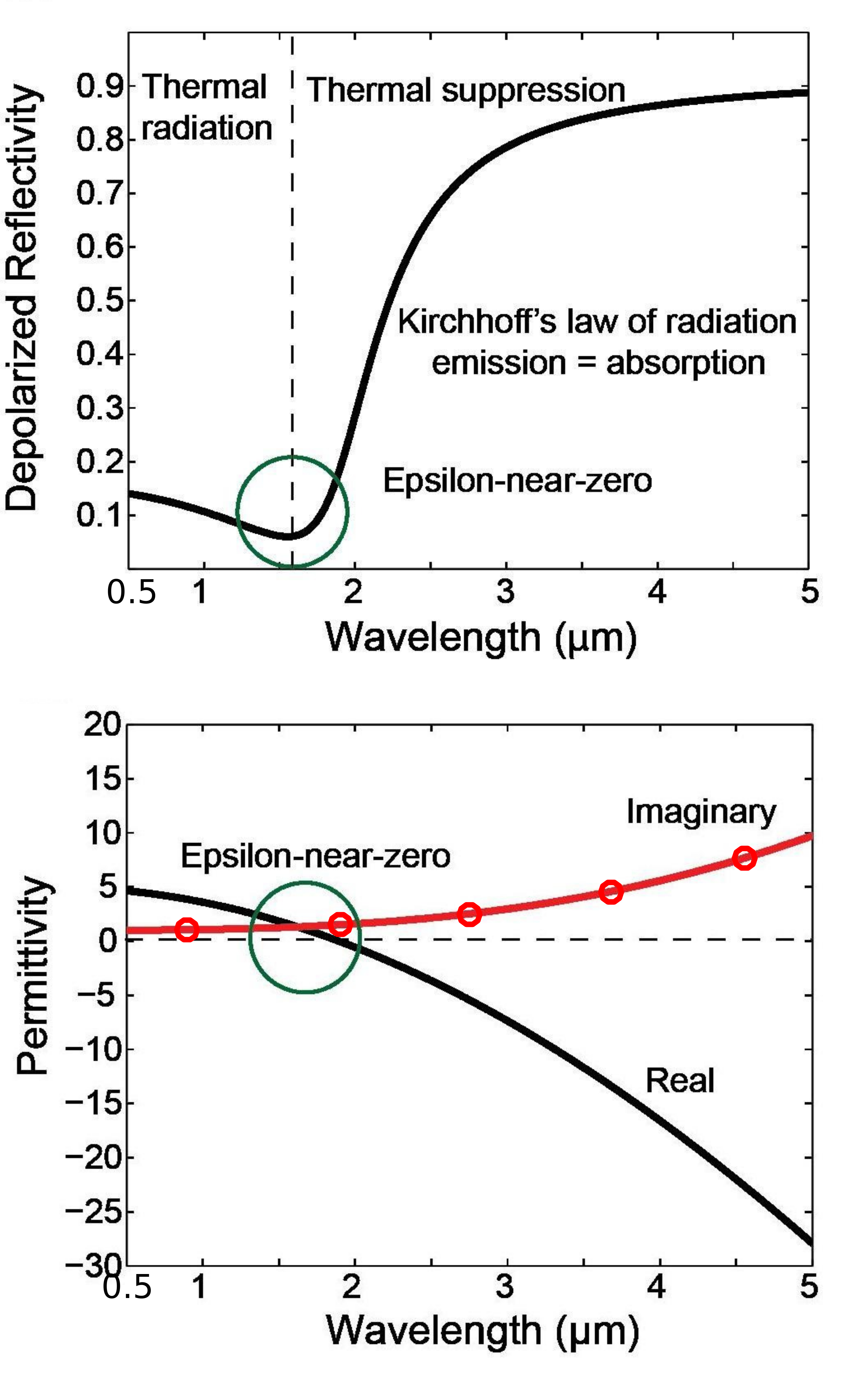}
\caption{\label{fig:Fig2}(a) Depolarized (averaged $s$ and $p$ polarized) normal reflectivity from plasmonic half space. High reflectivity beyond the epsilon-near-zero wavelength implies a low absorptivity/emissivity (thermal suppression of low energy sub-bandgap photons). Note, Kirchhoff’s laws require the far-field emissivity of the plasmonic half space to be its absorptivity.  (b) By shifting the plasma frequency (epsilon-vear-zero frequency) of the metal, the spectral content of the emitted radiation can be engineered to match the bandgap of a photovoltaic cell. This arises through the sharp change of the metal’s reflectivity at its plasma frequency. The sample Drude metal considered here is assumed to have background dielectric constant of 5, a plasma frequency of 0.66 eV and a loss parameter of 70 meV.}
\end{figure}

Selective thermal emitters for TPVs have to perform the important function of thermal suppression below the bandgap of the absorbing cell and simultaneously enhance emission just above the PV cell bandgap \cite{rephaeli2009absorber,lenert2014nanophotonic}. We design this feature by employing a thin film of Drude metal with engineered plasma frequency (ENZ frequency) and no surface patterning. The plasma frequency is the characteristic frequency above which metals lose their reflectivity and become transparent. Thus metals absorb above the plasma frequency and reflect light below the plasma frequency. Fig.~\ref{fig:Fig2}(a) and (b) show the correspondence between a sharp reflectivity change at normal incidence and the ENZ wavelength for a Drude metal. The high reflectivity leads to decreased optical absorption and suppresses the thermal emission by the Kirchoff’s law (absorptivity=emissivity).  One can  therefore achieve suppression of sub-bandgap thermal emission simply by tuning the plasma frequency in the near-infrared range close to the bandgap of the TPV cell. The design achieves enhancement in thermal emission in the transparency range of the metal tuned to be above the bandgap of the cell. Note, all metals are transparent and lossy above the ENZ frequency (plasma frequency) giving rise to large absorption and emission.

\section{High Temperature Material Properties}
We now outline the choices for high temperature metals that can act as a plamonic thermal emitter coating. Even though tungsten and tantalum have excellent thermal properties\cite{lenert2014nanophotonic}, the plasma frequency (ENZ frequency) cannot be tuned to the near-infared range to match the low energy bandgap of gallium antimonide or germanium TPV cells. We have performed detailed analysis of titanium nitride and zirconium nitride which are refractory metals with plasmonic response for TPV applications \cite{guler2014refractory,li2014refractory}. Our analysis (to be published later) indicate that their interband transitions and losses, especially in the near-infrared region are sub-optimal for both the far-field and the near-field plasmonic thermal emitter performance. They will need to be optimized to find use in TPV systems as a thin film plasmonic coating but can be used as nanoantennas \cite{guler2014refractory,li2014refractory}. In this work, we therefore focus on the use of aluminum doped zinc oxide (AZO) and gallium doped zinc oxide (GZO) with plasma frequency (ENZ frequency) tuned in the near-infrared for p-TECs \cite{pradhan2014extreme,kim2013optical}. 
In general, the optical response of materials at high temperatures poses a significant challenge for thermal emitter design.  We have utilized empirical models of tungsten and AZO/GZO available in literature to arrive at general conclusions of the real part and imaginary part of the response at high temperatures \cite{pradhan2014extreme,kim2013optical,krishnamoorthy2012topological,naik2011oxides,roberts1959optical,costescu2003thermal,schmid1998optical,larruquert2012self,liu2015enhanced}. Increasing temperature causes a rise in electron-phonon interactions and simultaneous reduction of collision time. This is manifested in a reduction of the bulk polarization response and increase in optical absorption. Consequently, the real part of the dielectric constant governing the polarization response is reduced and the imaginary part of the dielectric constant governing losses is increased. Thus the performance of plasmonic components is expected to be reduced at high temperature.  A detailed analysis of these empirical models will be published later.

\begin{figure}
\includegraphics[width=1\linewidth]{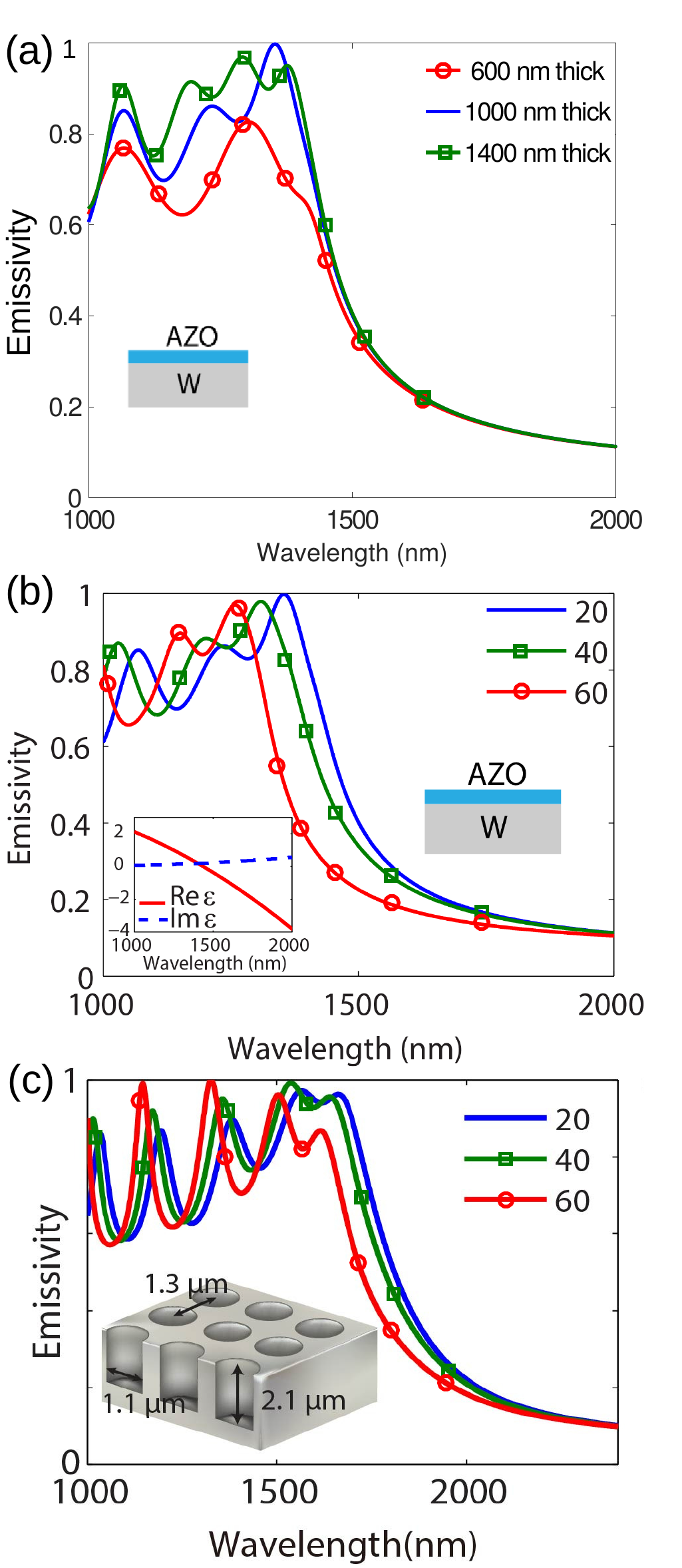}
\caption{\label{fig:Fig3} A single layer of AZO can suppress the long-wavelength emissivity, similar to a microfabricated tungsten photonic crystal. (a) Emissivity spectra of AZO thin film on tungsten substrate at 20$^\circ$ emission angle and varying AZO thickness.  (b) Emissivity spectra of 1000nm AZO film on W substrate at different emission angles.  Inset: Real and imaginary part of AZO dielectric permittivity. Beyond 1400nm the emissivity is suppressed by a large impedance mismatch with vacuum. (c) Emissivity of tungsten photonic crystals at different angles. Inset: Schematic of the PhC structure. By controlling the bandgap of photonic crystals, we can tune the emissivity cut-off match wavelength to the PV cell bandgap.}
\end{figure}

\begin{figure}
\includegraphics[width=1\linewidth]{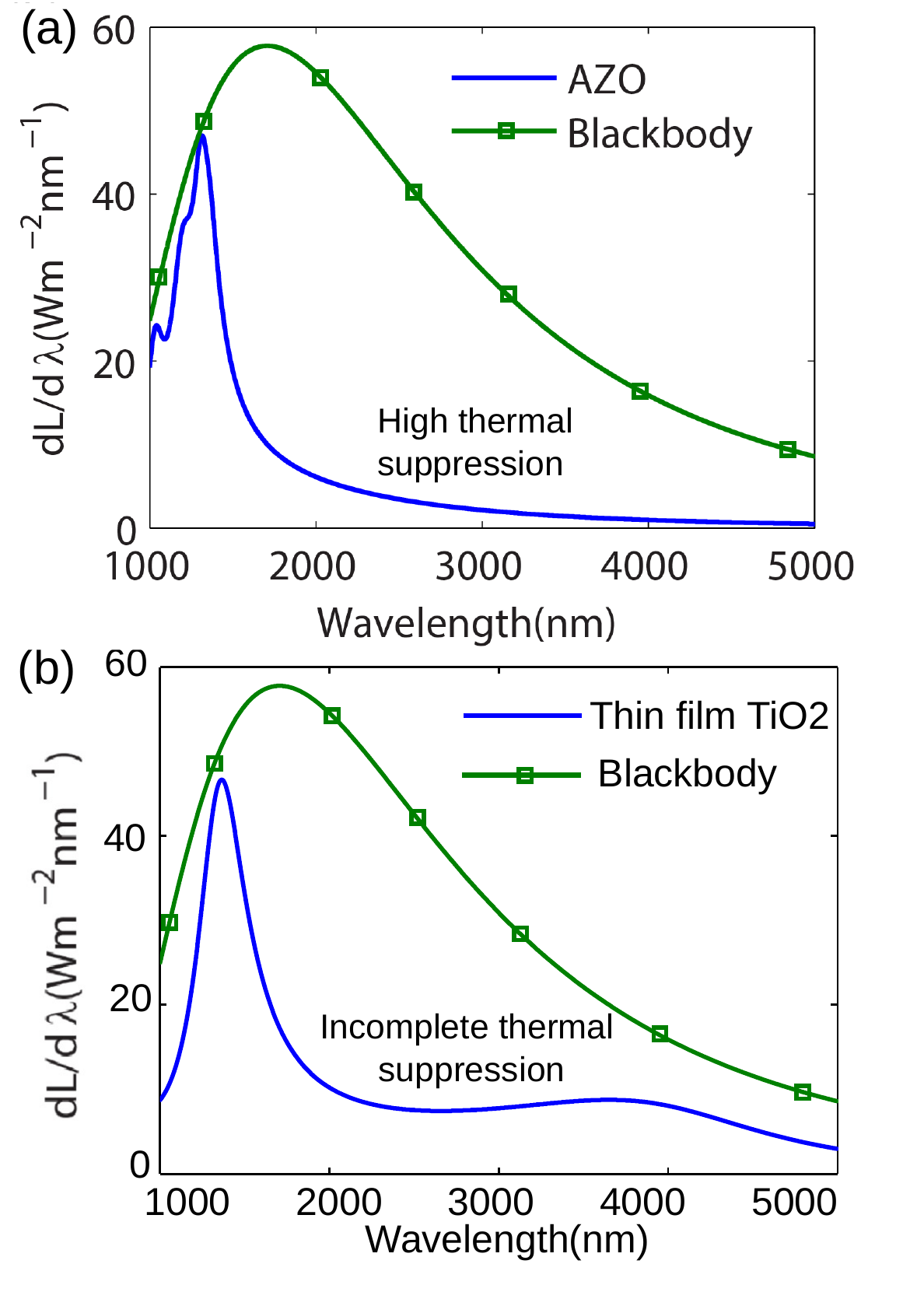}
\caption{\label{fig:Fig4} AZO thin-film outperforms TiO2 thin-film in supression of sub-bandgap thermal emission. (a)Spectral irradiance  ($dl/d\lambda$) of AZO single layer in comparison with a black body at 1700K (40 degrees emission angle). The selective spectral irradiance shows a large thermal suppression of the sub-bandgap photons and enhanced thermal emission below the ENZ wavelength (above the plasma energy). (b) Spectral irradiance of a single layer TiO2 anti-reflection coating design at 1700K and 40 degrees emission angle. The enhanced absorption in TiO2 near 1500~nm arises from the Fabry-Perot mode resonance so the peak emittance can be tuned by changing the thickness of the anti-reflection coating. However, the suppression of the sub-bandgap photons is incomplete in TiO2 layer, resulting in reduced conversion efficiency.}
\end{figure}

\section{High temperature thermal emission suppression}
\subsection{\label{sec:level2}Far-field thermal emission: thin-film AZO and other approaches}

We now provide the results of calculations using an AZO plasmonic thermal emitter coating. Using Kirchhoff’s laws \cite{greffet1998field}, we calculate the emissivity of an AZO film of thickness 600~nm, 1000~nm and 1400~nm, placed on a tungsten substrate (Fig. 3(a)). It can be observed that there is a sharp suppression in the emissivity above ENZ wavelength of 1400~nm, irrespective of the thickness of the AZO film. This is because the spectral thermal emission response is dominated by the material properties of the AZO and the thickness only effects the ripples in the high emissivity region. Fig.~3(b) shows the emissivity of a 1000~nm thick AZO film at different emission angles. The inset shows the real and imaginary part of AZO dielectric response. The emissivity performance at different emission angles displays a similar cut-off behavior, which is necessary for TPV applications. The p-TEC (Fig.~\ref{fig:Fig3}(b)) achieves an emissivity profile very close to a tungsten photonic crystal (PhC) design (Fig. 3(c)). However, note that there are fundamental differences in the approach to achieving thermal suppression. The PhC utilizes a structural resonance and interference effects whereas the p-TEC uses an engineered material response. A simple thin-film coating with engineered material response matches the performance of a microfabricated photonic crystal.

Thin-film p-TEC with engineered material response also outperforms thin-film anti-reflection coatings. Fig.~\ref{fig:Fig4} compares the  performance of an anti-reflection coating (AR) \cite{fraas2001antireflection} as a selective thermal emitter. Fig.~\ref{fig:Fig4}(a) shows the spectral irradiance from an AZO coated tungsten substrate at 1700K. On comparison with black body radiation, it can be clearly seen that the thermal radiation in the sub-bandgap region is suppressed. This sharp suppression at high wavelengths is because the emissivity of AZO coating drops above the epsilon-near-zero wavelength. The thickness of the the AZO film is 1000~nm. Fig.~\ref{fig:Fig4}(b) shows the thermal radiation from a tungsten coated with TiO2 AR coating. Enhancement in the thermal emission near 1500~nm wavelength in AR coating is due to the enhanced absorptivity arising from Fabry-Perot resonance mode in a TiO2 film of thickness 355~nm. The peak emittance in TiO2 can therefore be tuned by tuning the thickness of AR coating. This is in contrast to the AZO thin film, where the suppression in the emissivity is governed by the material response while the thickness has only a minor effect on the emissivity at transparent low-wavelength region of the spectrum. It can be seen that the even at sub-bandgap wavelengths, the thermal emission in TiO2 is considerably larger as compared to the AZO coating. The incomplete thermal suppression in AR coating degrades its efficiency performance.   

To provide a performance comparison in the limiting ideal case of these various emitter designs for TPV applications, we apply the Schockley-Queisser detailed balance analysis \cite{rephaeli2009absorber}. In performing this calculation we assume the photovoltaic cell in all cases to be a perfectly absorbing, ideal pn-junction, with a bandgap of 2250 nm for the photonic crystal emitter and 1700 nm for the AZO plasmonic coating and titanium dioxide anti-reflection  coating designs. We also assume that the emitter and photovoltaic cell have the same flat area, and that no absorbing stage is present.  Under these conditions, the conversion efficiency of radiated thermal energy into electrical energy in AZO, photonic crystal and TiO2 coating is shown in Table~\ref{Tab1}. The efficiency of a simple AZO thin-film is less then, but comparable with the photonic crystal spectrally selective emitter, while it outperforms the TiO2 thin film. The poor efficiency of TiO2 is because of incomplete thermal suppression at higher wavelengths.

\begin{table}[H]
\caption{Comparison of conversion efficiency in AZO, photonic crystal and TiO2 AR coating.}
\label{Tab1}
\centering
\begin{tabular}{ |c| c| c| c|}
\hline
Temperature     & AZO thin-film   & Photonic crystal & TiO2 \\ \hline 
1700K & 31.9 \%  & 36.7 \%    & 19.6 \% \\ \hline
1300K & 19.4 \%  & 29.1 \%    & 9.2  \% \\ \hline
\end{tabular}
\end{table}

\subsection{ Narrow-band near-field thermal emission}

\begin{figure}[h]
\includegraphics[width=1\linewidth]{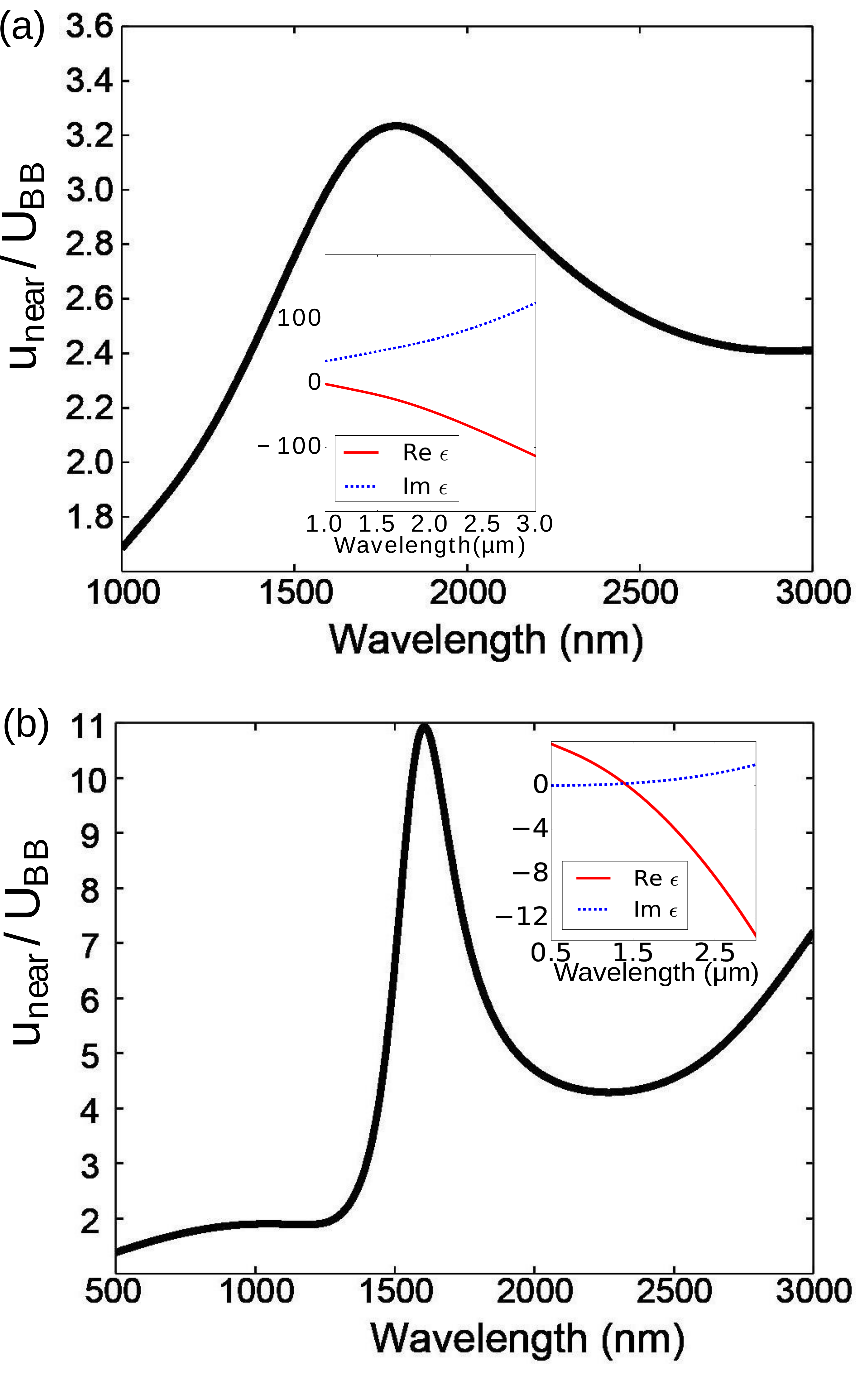}
\caption{\label{fig:Fig5}Near-field energy density at a distance of 100nm above (a) a 20nm tungsten film on a sapphire substrate, and (b) a 55nm gallium-doped-zinc-oxide film [18] on a boron carbide substrate [22], relative to that of a black body. The plasmonic thermal emitter coating shows a near-field enhancement by a factor of three over the thin film tungsten. Inset in (a) and (b) shows the permittivity of Tungsten and GZO respectively.}
\end{figure}

A fundamental advantage of the p-TEC design is that it can simultaneously be used as a near-field emitter and can exhibit narrowband near-field thermal emission. This arises due to thin film surface plasmon polaritons excited at high temperatures which lead to narrowband thermal emission with potential applications for near-field TPV. While fundamentally challenging to implement, near-field TPVs promises to achieve high efficiency of conversion with high current densities\cite{basu2009maximum}. This is because near-field thermal energy transfer mediated by evanescent waves can exceed the far-field black body limit.

In Fig.~\ref{fig:Fig5}(a), we plot the energy density in the near-field of a conventional emitter used in near-field TPV designs consisting of a thin film of tungsten. The energy density is normalized to that of a black body in the far-field and calculated using Rytov’s fluctuational electrodynamics\cite{nagasaka2008zhuomin,liu2015enhanced}. The energy density at a distance $z$ and frequency $\omega$  is \cite{joulain2007radiative}
\begin{equation*}
\begin{split}
u(z,\omega,T)&=\frac{U_{BB}(\omega,T)}{2} \!\! \left \{ \!\! \int_{0}^{k_0} \!\! \frac{k_\rho dk_\rho}{k_0|k_{1z}|}\frac{(1-|r^s|^2)+(1-|r^p|^2)}{2} \right.  \\
& \left. +\int_{k_0}^{\infty}\frac{k^3_\rho dk_\rho}{k^3_0 |k_{1z}|}e^{-2Im(k_{1z})z}(Im(r^s)+Im(r^p))\right\}
\end{split}
\end{equation*}
where  $T$ denotes the temperature.  $k_\rho=\sqrt{k^2_x+k^2_y}$, $k_0=\omega/c$  , $k_{1z}=\sqrt{k^2_0-k^2_\rho}$  (such that $Im k_{1z}>0$), $r^s$and $r^p$ are the Fresnel reflection coefficients from the thin films for (s) and (p) polarized light respectively and “$Im$” denotes the imaginary part. $U_{BB}(\omega,T)$  is the black body emission spectrum. The near-field energy density enhancement is due to the excitation of weak and lossy surface waves in highly absorptive tungsten (above its plasma wavelength of $\approx 936~nm$). Due to the lossy nature of tungsten at near-infrared wavelengths \cite{roberts1959optical}, the enhancement is low, with a broad spectrum. This poor spectral performance is a significant limiting factor for TPVs. 

Fig. 5(b) shows the near-field energy density near a thin film of Gallium doped Zinc-oxide (GZO). The plasma frequency (ENZ frequency) of doped zinc-oxide can be tuned to be in the near-infrared spectral range, matched to the bandgap of a TPV cell (GaSb). The spectrally selective nature along with the larger enhancement of near-field energy density, evident in Fig.5(b), is due to the excitation of surface plasmons. By controlling the film thickness and ENZ permittivity of the GZO film, the spectral content of near-field enhancement can tuned to match the high efficiency window of energies directly above the band-gap of a PV cell. For comparison, the thickness of tungsten and gallium doped-zinc-oxide are selected such that the peak in energy density enhancement is optimized at the same frequency. The plasmonic thermal emitter coating (p-TEC) outperforms the thin film of tungsten and can be used in both near-field and far-field TPV.

\begin{figure}[h]
\centering
\includegraphics[width=1\linewidth]{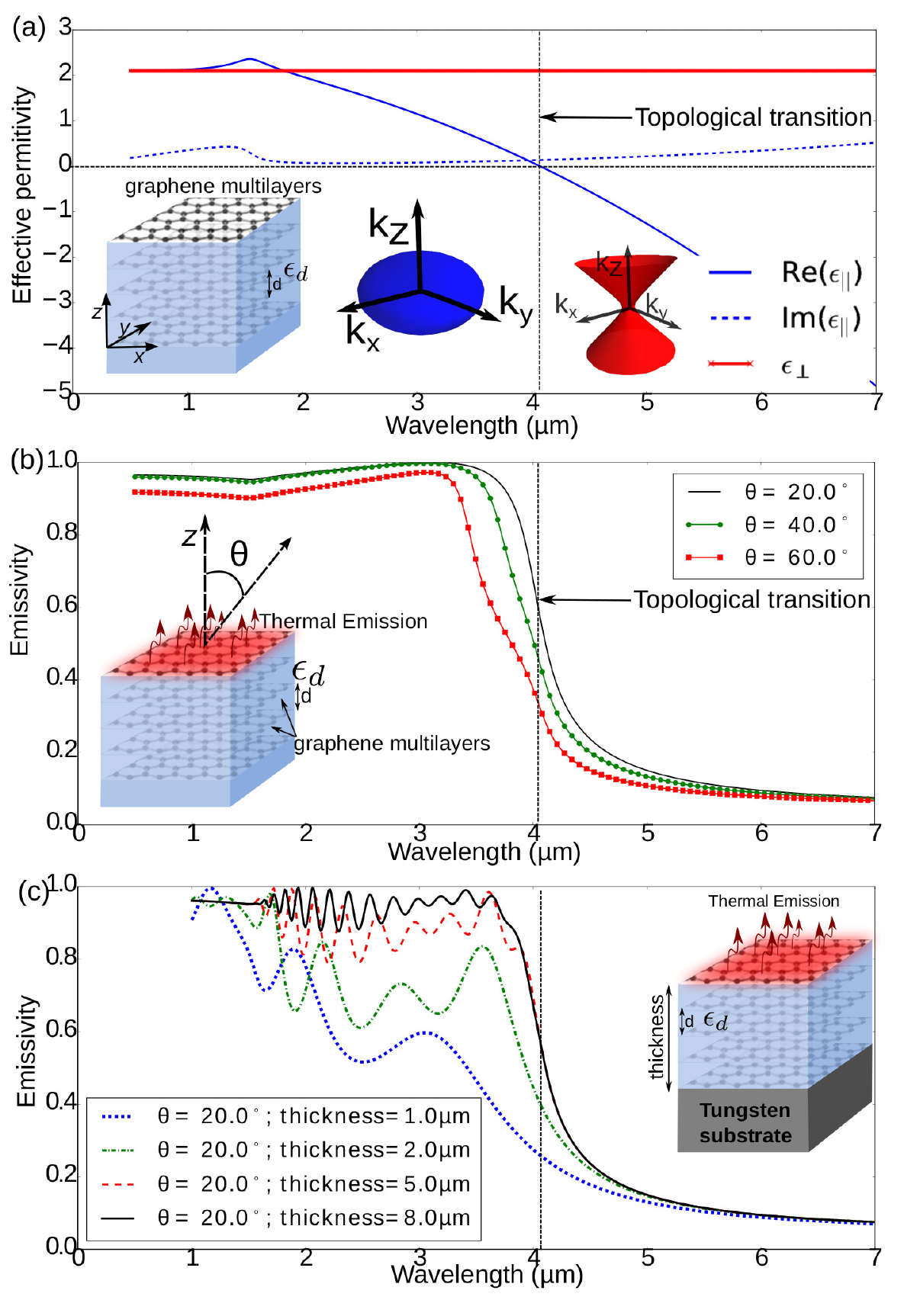}
\caption{Far-field emissivity of Graphene multilayer. (a) shows the effective permittivity of graphene multilayer substrate when graphene layers are separated by a dielectric of permittivity 2.1 and thickness 10 nm. The Fermi energy of the graphene is $E_F$ = 0.4 eV. The effective permittivity perpendicular to the graphene layers remains unaffected while the parallel component drops as the excitation wavelength increases. The point where $Re(\epsilon_{||})$ crosses the zero is the topological transition point. (b) shows the emissivity of bulk graphene-multilayer at three different angles (c) shows the emissivity of varying finite thickness of graphene-multilayer deposited on a Tungsten substrate.}
\label{fig:Graphene_far_field_emission}
\end{figure}

\begin{figure}
\centering
\includegraphics[width=1\linewidth]{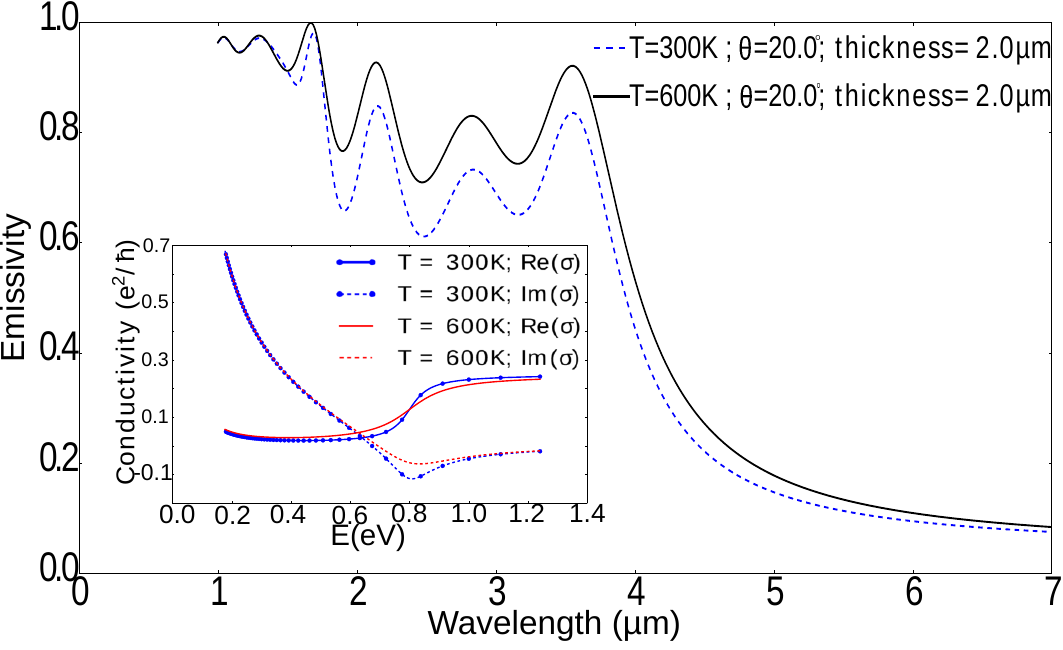}
\caption{Effect of temperature on the far-field emissivity response of the graphene-multilayer structure. The conductivity of the graphene at 300K and 600K is shown in the inset.}
\label{fig:Graphene_far-field_vary_Temp_thickness=2micrometer}
\end{figure}

\begin{figure}
\centering
\includegraphics[width =1\linewidth]{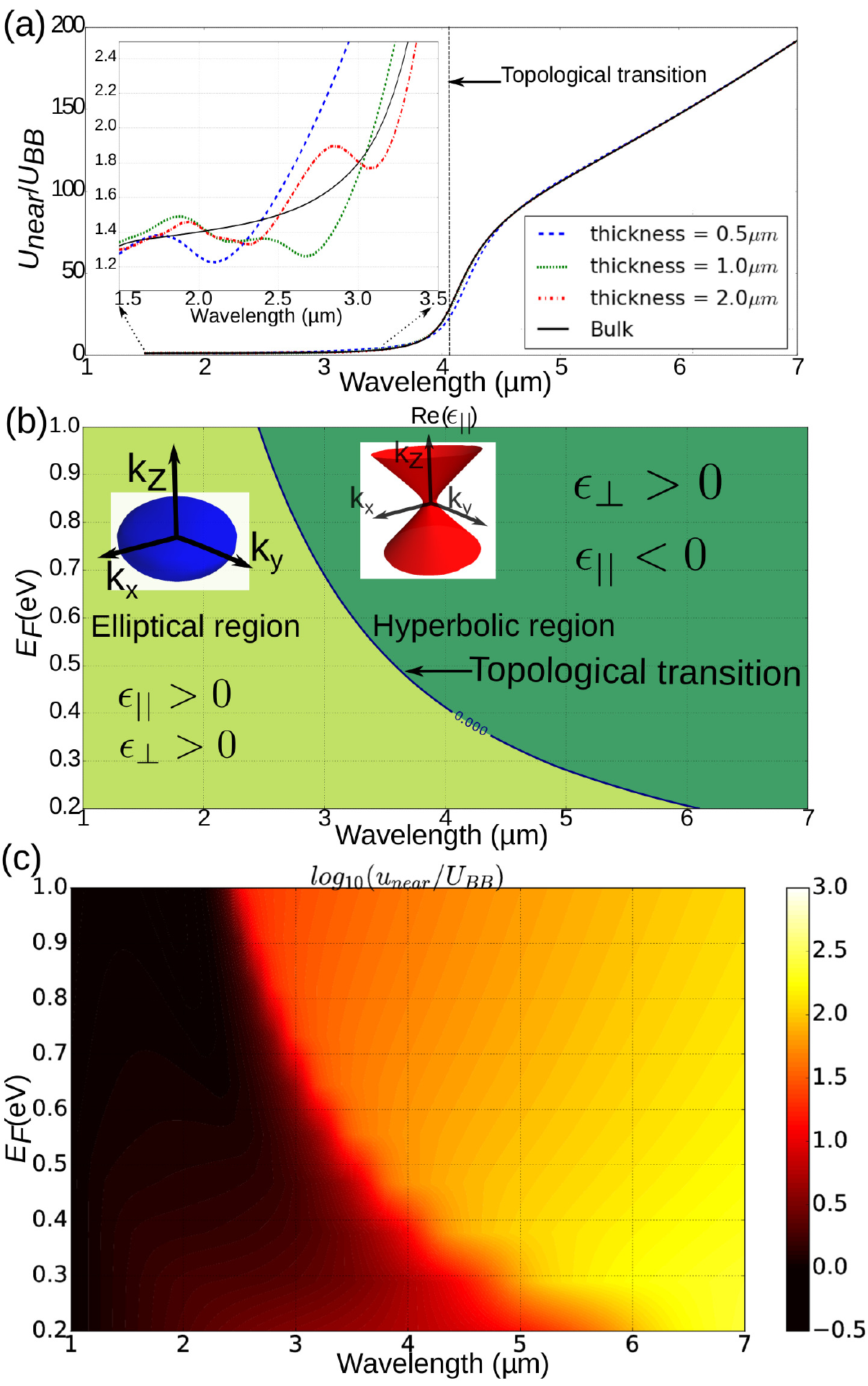}
\caption{ (a) Near-field energy density near the graphene-multilayer substrate at a distance of 100 nm above the substrate. Inset shows that the effect of thickness on the energy density is negligible as compared to enhancement due to bulk topological states. (b) shows the sign of $Re(\epsilon_{||})$  as a function of $E_F$ and excitation wavelength. Topological transition from elliptic to hyperbolic substrate occurs on the line $Re(\epsilon_{||}) =0 $. (c) shows the near-field energy density as function of $E_F$ and excitation wavelength. It can be seen that hyperbolic region of the graphene multilayer enhances the near-surface energy states. The thermal emission is computed at $600^\circ K$.}
\label{fig:Graphene_near_field_emission}
\end{figure}

\section{Thermal topological transitions in graphene multilayers}

Thermal emission can also be engineered by controlling the topological transition in a graphene-multilayer structure. A schematic of the graphene multilayers stacked between dielectric slabs of $\epsilon_d=2.1$ and thickness $d= 10$nm  is shown in inset of Fig.~\ref{fig:Graphene_far_field_emission}(a). The layers are stacked along the $z$ direction. A multilayer structure thus formed has an anisotropic dielectric tensor, whose perpendicular component (along the $z$-direction) of the permittivity ($\epsilon_\perp$) is given by the dielectric $\epsilon_d$, while the parallel component ($\epsilon_{||}$, parallel to the layers in $x-y$~plane) is governed a combination of three factors: (i) the complex conductivity of graphene, (ii) the thickness of the separating substrate ($d$) and (iii) the dielectric $\epsilon_d$. The equations for computing the dielectric tensor of graphene metamaterial is presented in Appendix. Fig.~\ref{fig:Graphene_far_field_emission}(a) shows the parallel component (in $x-y$ plane) and the perpendicular component (along $z$-direction) of the dielectric tensor, when Fermi energy of each graphene layer is $E_F=0.4$~eV. It can be seen that the parallel component of permittivity changes sign from positive to negative around 4.05~$\mu$m, triggering a topological transition from elliptical to hyperbolic iso-frequency surface, as shown in the inset. At the topological transition wavelength, the emissivity of the structure  drops sharply due to large miss-match in the hyperbolic topology of the graphene-multilayer substrate and the elliptical topology of the free-space. Fig.~\ref{fig:Graphene_far_field_emission}(b) shows the suppression of emissivity at different angles from bulk graphene metamaterial. For thermal emission applications the graphene metamaterial substrate will need to be deposited on a substrate, such as tungsten. Fig.~\ref{fig:Graphene_far_field_emission}(c) shows the emission characteristics of graphene metamaterial deposited on tungsten substrate for varying thickness of the graphene-multilayer. It can be seen that as the thickness increases beyond 5~$\mu$m, its emission characteristics approach those of the bulk graphene. These computations are done for a Fermi-energy of 0.4~eV. 

We would like to point out that the wavelength at which topological transition in graphene metamaterial occurs can be controlled by the dielectric thickness $d$ as well as the Fermi energy of the individual graphene sheets. This gives us an additional degree of freedom in controlling the topological transition wavelength. The graphene metametrial is also suited for high temperature applications as its doping concentration and complex conductivity is does not vary substantially with temperature \cite{fang2015temperature}. Fig.~\ref{fig:Graphene_far-field_vary_Temp_thickness=2micrometer} shows the variation in emissivity at temperature T=300K and T=600K. The change in complex conductivity of individual graphene sheets at the two temperatures is shown as inset. It can be seen that the even at a higher temperature, the emissivity of the graphene metamaterial is suppressed at higher wavelengths. 

The topological transition in graphene-multilayers also results in enhanced near-field thermal emission. Graphene multilayers support unbounded bulk hyperbolic states which increase the local density of states in vacuum, near the surface. This increased local density of states results in enhanced near-field energy density and near-field thermal emission. Fig.~\ref{fig:Graphene_near_field_emission}(a) shows the large enhancement in the near-field energy density (normalized to that of black body), in the hyperbolic region of the spectrum. It can be observed that there is a sharp increase in energy density at the topological  transition wavelength. The near-field energy is computed for increasing thickness of graphene-multilayer on a tungsten substrate. It can be seen in the zoomed-in inset that the finite thickness has small effect on the near-field as compared to large enhancement triggered by topological transition. The near-field energy is computed at a distance of 100 nm from the substrate. The topological transition and hence the near-field emission can be tuned by controlling the Fermi energy of the graphene layers. The contour on which the topological transition occurs as function of wavelength and Fermi-energy is shown in Fig.~\ref{fig:Graphene_near_field_emission}(b). The corresponding enhancement in the thermal emission at $600$K is shown in Fig.~\ref{fig:Graphene_near_field_emission}(c). It is evident that the enhancement in near-field energy density is triggered at the topological transition for all values of Fermi energy.

\section{Summary}
In conclusion, we have introduced the concept of a plasmonic thermal emitter coating. It functions on the basis of reflectivity change of metallic thin films near the epsilon-near-zero frequency. Our approach shows superior performance as compared to anti-reflection coatings and is easier to fabricate than photonic crystals which require 2D surface patterning of tungsten. We have shown that it achieves narrowband thermal emission in the near-field as compared to tungsten. Developments in high temperature plasmonics can make our thin film design a viable large area technology for thermal emitters. We have also shown that the thermal topological transitions in graphene multilayers can lead to tunable spectrally selective thermal emission.

\section*{Acknowledgements}

We acknowledge funding from Alberta Innovates Technology Futures, NSERC, Helmholtz Alberta Initiative and NSF EFRI NEWLAW.

\appendix

\section{Thermal emission from Fluctuation Dissipation Theorem }
Any material consists of a large ensemble of fluctuating currents which are the source of thermal radiation. The Poynting vector of the thermal radiation from a from a material is governed by the Fluctuation Dissipation Theorem. The thermal radiation from a material half-space, in the $z$-direction, is given by \cite{joulain2007radiative},

\begin{equation}
\langle S_z(\vec{r},\omega)\rangle = \frac{\omega^2}{4\pi^2 c^2}\frac{\hbar\omega}{e^{\hbar\omega/(K_bT)}-1}\int_{\Omega=2\pi} \!\!\!\! A_{emi} \frac{\cos\theta d\Omega}{2\pi}
\end{equation}
where $\theta$ is the angle from the normal ($z$-axis) to the interface, and $A_{emi}$ is the emissivity of the material, given by,

\begin{equation}
A_{emi}=\frac{\left(1-|r_{12}^s|^2+1-|r_{12}^p|^2\right)}{2}
\end{equation}
where, $r_{12}^s$ and $r_{12}^p$ is the reflection coefficient of $\hat s$ and $\hat p$ polarized wave at the interface of the material and vacuum. 

We numerically compute the reflection coefficient at the vacuum/thin-film interface for a $\hat s$ and $\hat p$ polarized wave. These reflection coefficients as a function of frequency and incident angles are then used to compute the far-field and near-field thermal emission. 

\section{Effective permittivity of graphene-multilayers }

The complex conductivity of graphene is given by \cite{wunsch2006dynamical,hwang2007dielectric},

\begin{equation}
\sigma\left(\omega,E_F\right) =\sigma_{intra}+\sigma_{inter}
\end{equation}

where, 

\begin{equation}
\sigma_{intra} = \frac{2i}{\pi}\left(\frac{K_bT}{(\omega+i\tau^{-1})\hbar}\right)\ln\left( 2\cosh\left(\frac{E_F}{2T K_b}\right)\right)
\end{equation}

and 

\begin{equation}
\begin{split}
\sigma_{inter} &= \frac{1}{4}\left[\frac{1}{2} + \frac{1}{\pi}\tan^{-1}\left(\frac{\omega \hbar -2E_F}{2T K_b}\right) \right. \\ 
&-\left. \frac{i}{2\pi}\ln\left(\frac{\omega \hbar +2 E_F}{(\omega\hbar -2 E_F)^2 + (2 T K_B)^2}\right) \right]
\end{split}
\end{equation}

Relaxation time $\tau=6\times10^{-14}$s, $T$ is temperature in Kelvin. Where $\sigma_{intra}$ and $\sigma_{inter}$ corresponds to intra-band and inter-band conductivity, respectively; $K_B$ is the Boltzmann constant, and $E_F$ is the Fermi Energy.

When the graphene multilayers are stacked between material of dielectric constant $\epsilon_d$ and thickness $d$ (as shown in inset of Fig.~\ref{fig:Graphene_far_field_emission}(a)), the resultant material is anisotropic, with dielectric tensor,

\begin{equation}
\overleftrightarrow{\epsilon} = \begin{bmatrix}
\epsilon_{||} & 0 & 0 \\
0 & \epsilon_{||} & 0 \\
0 & 0 & \epsilon_\perp
\end{bmatrix}
\label{eq:appendix_b4}
\end{equation}
where $\epsilon_{||}$ and $\epsilon_\perp$ are the $x-y$ and $z$ component of the effective dielectric, respectively \cite{chang2016realization}.

\begin{eqnarray}
\label{Eq:epsilon_parallel} 
\epsilon_{||} & = & \epsilon_d + \frac{i\sigma\left(\omega,E_F\right)}{\omega \epsilon_0 d}\\
\label{Eq:epsilon_perp}
\epsilon_\perp & = & \epsilon_d
\end{eqnarray}

For computation of thermal emission from graphene metamaterial, the  reflection coefficients are computed for film of anisotropic dielectric tensor given by equation~(\ref{eq:appendix_b4}).



\bibliography{ThermalApp.bib}

\end{document}